\documentclass[11pt]{article}
\usepackage{geometry}                % See geometry.pdf to learn the layout options. There are lots.
\usepackage{graphicx}
\usepackage{amssymb}
\usepackage{epstopdf}
\usepackage{amsmath,amsfonts,amssymb,amsthm}
\usepackage{enumerate}
\usepackage{epsfig}
\usepackage{scalefnt}

\usepackage{setspace}
\title{Parameter Selection Methods in Inverse Problem Formulation}

\author{ H. T. Banks$^{1}$ and Ariel Cintr\'{o}n-Arias$^{1,2}$\\\\
Center for Research in Scientific Computation \\
Center for Quantitative Sciences in Biomedicine \\
Department of Mathematics\\
$^1$North Carolina State University \\
Raleigh, NC 27695-8212\\
and\\
$^2$Department of Mathematics and Statistics\\
East Tennessee State University\\
Johnson City, TN  37614-0663}

\date{May 28, 2010}                                           % Activate to display a given date or no date

\begin{document}
\maketitle
%\vspace{.3in}

\vspace{.3in}
\begin{abstract}
\noindent We discuss methods for {\em a priori} selection of parameters to be estimated in inverse problem formulations (such as Maximum Likelihood, Ordinary and Generalized Least Squares) for dynamical systems with numerous state variables and an even larger number of parameters. We illustrate the ideas with an in-host model for HIV dynamics which has been successfully validated with clinical data and used for prediction.
\end{abstract}

\vspace{1in}
{\bf Key Words:} Parameter selection, inverse problems, sensitivity, Fisher Information Matrix, HIV models.

%%\tableofcontents
\clearpage
\section{Introduction}
There are many topics of great importance and interest in the areas of modeling and inverse problems which are properly viewed as essential in the use of mathematics and statistics in scientific inquiries. A brief, noninclusive list of topics include the use of traditional sensitivity functions (TSF) and generalized sensitivity functions (GSF) in experimental design (what type and how much data is needed, where/when to take observations) \cite{BDE,BDEK,BEG,Kappel,TC}, choice of mathematical models and their parameterizations (verification, validation, model selection and model comparison techniques) \cite{BDSS,BF1,BF2,bed,boz1,boz2,BA1,BA2,hur}, choice of statistical models (observation process and sampling errors, residual plots for statistical model verification, use  of asymptotic theory and bootstrapping for computation of standard errors, confidence intervals) \cite{BDSS,BHR,DG,ET,SeWi,ST}, choice of cost functionals (MLE, OLS, WLS, GLS, etc.,) \cite{BDSS,DG}, as well as parameter identifiability and selectivity. There is extensive literature on each of these topics and many have been treated in surveys in one form or another (\cite{DG} is an excellent monograph with many references on the statistically related topics) or in earlier lecture notes \cite{BDSS}.

We discuss here an enduring major problem: selection of which model parameters can be readily and reliably (with quantifiable uncertainty bounds)  estimated in an inverse problem formulation. This is especially important in many areas of biological modeling where often one has large dynamical systems (many state variables), an even larger number of unknown parameters to be estimated and a paucity of longitudinal time observations or data points. As biological and physiological models (at the cellular, biochemical pathway or whole organism level) become more sophisticated (motivated by increasingly detailed understanding - or lack thereof - of mechanisms), it is becoming quite common to have large systems (10-20 or more differential equations), with a plethora of parameters (25-100) but only a limited number (50-100 or fewer) of data points per individual organism. For example, we find models for the cardiovascular system \cite[Chapter 1]{Kappel} (where the model has 16 state variables and 22 parameters) and \cite[Chapter 6]{Ottesen} (where the model has 22 states and 55 parameters), immunology \cite{nelson} (8 states, 24 parameters), metabolic pathways \cite{Engl} (8 states, 35 parameters) and HIV progression \cite{BDHKR,jones} 
(8 \& 6 states, 11 \& 8 parameters, respectively).  Fortunately, there is a growing recent effort among scientists to develop quantitative methods based on sensitivity, information matrices and other statistical constructs (see for example \cite{BDE,BDEK,BEG,burth,ac09,fink,fink2,wu}) to aid in identification or parameter estimation formulations. We discuss here one approach using sensitivity matrices and asymptotic standard errors as a basis for our developments. To illustrate our discussions, we will use a recently developed in-host model for HIV dynamics which has been successfully validated with clinical data and used for prediction \cite{adams07,BDHKR}.

The topic of system and parameter identifiability is actually an old one.  In the
context of parameter determination from system observations or
output it is at least forty years old and has received much attention in
the peak years of linear system and control theory in the
investigation of observability, controllability and detectability
\cite{AE,bellams,BeKa,Eykhoff,GW,Kalman,MehraLain,reid77,SageMelsa}.
These early investigations and results were focused primarily on
engineering applications, although much interest in other areas (e.g.,
oceanography, biology) has prompted more recent inquiries for both
linear and nonlinear dynamical systems
\cite{anh06,BS,cob80,evans05,holm82,navon,white01,wu,xia03,yue08}.

\subsection{A Mathematical Model for HIV Progression with Treatment Interruption}\label{modelsection}

We summarize and use as an illustrative example one of the many dynamic models for HIV progression found in an extensive literature (e.g., see \cite{adamsthesis,brian1,brian2,adams07,BDHKR,Bon,callaway,NowakBangham,PerelsonReview,WodarzNowak} and the many references therein). For our example model, the dynamics of in-host HIV are described by the interactions between uninfected and infected type 1 target cells ($T_1$ and $T_1^*$) (CD4$^{+}$ T-cells), uninfected and infected type 2 target cells ($T_2$ and $T_2^*$) (such as macrophages or memory cells, etc.), infectious free virus $V_I$, and immune response $E$ (cytotoxic T-lymphocytes CD8$^+$) to the infection. This model, which was developed and studied in \cite{adamsthesis,adams07} and later extended in subsequent efforts (e.g., see \cite{BDHKR}), is essentially one suggested in
 \cite{callaway}, but includes an immune response compartment and dynamics as in 
\cite{Bon}. The model equations are given by

\begin{equation}
\begin{array}{l}
\dot{T}_1 = \lambda_1 - d_1 T_1 - \left(1 - \bar{\epsilon}_1 (t)\right) k_1 {V_I} T_1  \\
\dot{T}_2 = \lambda_2 - d_2 T_2 - (1- f \bar{\epsilon}_1 (t)) k_2 {V_I} T_2  \\
\dot{T}_1^* = (1 - \bar{\epsilon}_1 (t) )k_1 {V_I} T_1 - \delta {T_1^*} - m_1 {E} {T_1^*} \\
\dot{T}_2^* = (1 - f \bar{\epsilon}_1 (t))k_2 {V_I} T_2 - \delta {T_2^*} - m_2 {E} {T_2^*} \\
\dot{V}_I = (1 - \bar{\epsilon}_2 (t)) 10^3 N_T \delta ({T_1^*} + {T_2^*}) - c {V_I}   \\
           \hspace{0.5in} - (1 - \bar{\epsilon}_1 (t)) 10^3 k_1 T_1 V_I - (1 - f\bar{\epsilon}_1 (t)) 10^3 k_2 T_2 V_I \\
\dot{E} = \lambda_E + \frac{b_E ({T_1^*} + {T_2^*})}{({T_1^*} + {T_2^*}) + K_b}{E} - \frac{d_E ({T_1^*} + {T_2^*})}{({T_1^*} + {T_2^*}) + K_d}{E} - \delta_E {E}, %
\label{EQN_E_dynamics}
\end{array}
\end{equation}
together with an initial condition vector $\left( T_1(0), T_1^*(0), T_2(0), T_2^*(0), V_I(0), E(0) \right)^T.$

The differences in infection rates and treatment
efficacy help create a low, but non-zero, infected cell steady
state for $T_2^*$, which is compatible with the idea that
macrophages or memory cells may be an important source of virus after T-cell
depletion.  The populations of uninfected target cells $T_1$ and
$T_2$ may have different source rates $\lambda_i$ and natural
death rates $d_i$. The time-dependent treatment factors
$\bar{\epsilon}_1(t) = \epsilon_1 u(t)$ and $\bar{\epsilon}_2(t) =
\epsilon_2 u(t)$ represent the effective treatment impact of a reverse transcriptase
inhibitor (RTI) (that blocks new infections) and a protease inhibitor (PI) (which causes
infected cells to produce non-infectious virus), respectively. The RTI is potentially more
effective in population 1 ($T_1, T_1^*$) than in population 2
($T_2, T_2^*$), where the efficacy is $f \bar{\epsilon}_1$, with
$f \in [0,1]$. The relative effectiveness of RTIs is
modeled by $\epsilon_1$ and that of PIs by $\epsilon_2$,
while the time-dependent treatment function $0 \leq u(t) \leq 1$
represents therapy levels drug level, with $u(t) = 0$ for fully off and
$u(t) = 1$, for fully on. Although HIV treatment is nearly always administered as combination
therapy, the model allows the possibility of monotherapy, even
for a limited period of time, implemented by
considering separate treatment functions $u_1(t), u_2(t)$ in the treatment factors.

As in \cite{adamsthesis,adams07}, for our numerical investigations we consider a log-transformed and reduced version of the model. This transformation is frequently used in the HIV modeling literature because of the large differences in orders of magnitude in state values in the model and the data and to guarantee non-negative state values as well as because of certain probabilistic considerations (for further discussions see \cite{adams07}).   This results in the nonlinear system of differential equations
  \begin{eqnarray}
    	\frac{d x_1}{dt}&=& \frac{10^{-x_1}}{\ln(10)}\left( \lambda_1-d_1 10^{x_1}-(1-\bar\varepsilon_1(t))k_1 10^{x_5} 10^{x_1}\right) \label{x1eqn}\\\
    	\frac{d x_2}{dt}&=&\frac{10^{-x_2}}{\ln(10)}\left( (1-\bar\varepsilon_1(t))k_110^{x_5}10^{x_1}-\delta 10^{x_2}-m_1 10^{x_6}10^{x_2}\right)\\
    	\frac{d x_3}{dt}&=& \frac{10^{-x_3}}{\ln(10)}\left(\lambda_2 -d_210^{x_3}-(1-f\bar\varepsilon_1(t))k_2 10^{x_5}10^{x_3}\right)\\
    	\frac{d x_4}{dt}&=&\frac{10^{-x_4}}{\ln(10)}\left((1-f\bar\varepsilon_1(t))k_210^{x_5}10^{x_3} -\delta 10^{x_4} -m_2 10^{x_6}10^{x_4}\right)\\
   \nonumber \\
    	\frac{d x_5}{dt}&=&\frac{10^{-x_5}}{\ln(10)} ( (1-\bar\varepsilon_2(t))10^3N_T\delta(10^{x_2}+10^{x_4})-c10^{x_5}- \nonumber\\
    	&& \quad\quad\quad \quad(1-\bar\varepsilon_1(t))\rho_1 10^3 k_1 10^{x_1}10^{x_5} -(1-f\bar\varepsilon_1(t))\rho_2 10^3k_210^{x_3}10^{x_5})\\
    	\frac{d x_6}{dt}&=&\frac{10^{-x_6}}{\ln(10)}\left(\lambda_E+\frac{b_E(10^{x_2}+10^{x_4})}{(10^{x_2}+10^{x_4})+K_b}10^{x_6}
    	-\frac{d_E(10^{x_2}+10^{x_4})}{(10^{x_2}+10^{x_4})+K_d}10^{x_6}-\delta_E 10^{x_6}\right) \label{x6eqn},
  \end{eqnarray}
where the changes of variables are defined by

	\begin{equation}
%		\begin{array}{c}
			T_1=10^{x_1},\\
			T_1^*=10^{x_2},\\
			T_2=10^{x_3},\\
			T_2^*=10^{x_4},\\
			V_I=10^{x_5},\\
			E=10^{x_6}.
%		\end{array}
	\end{equation}	
We note that this model contains six state variables and twenty-two (in general, unknown) system parameters given by
\[ \theta_2= (\lambda_1,d_1,\epsilon_1,k_1,\lambda_2,d_2,f,k_2,\delta,m_1,m_2,
	\epsilon_2,N_T,c,\rho_1,\rho_2,\lambda_E,b_E,K_b,d_E,K_d,\delta_E).
\]

A list of the model parameters along with units of these model parameters are given below in Table \ref{tabpars}.

The initial conditions for equations (\ref{x1eqn})--(\ref{x6eqn}) are denoted by $x_i(t_0)=x_i^0$, for $i=1,\dots,6$.  We will also consider the initial conditions as unknowns and we use the following notation for the vector of parameters and initial conditions:
%%\begin{equation}
\[
\theta=(\theta_1,\theta_2)\]
where
\[	
\theta_1=(x_1^0,x_2^0,x_3^0,x_4^0,x_5^0,x_6^0)^T.
\]
%%\end{equation}

\begin{table}[h]
	\caption{Parameters for the HIV model.}
	\begin{center}
		\begin{tabular}{ccl} \hline\hline
		Parameter & Units & Description \\ \hline\hline
		$\lambda_1$&$\frac{\mbox{cells}}{\mbox{ml} \ \mbox{day}}$& Target cell type 1 production rate\\
		$d_1$&$\frac{1}{\mbox{day}}$& Target cell type 1 death rate\\
$\epsilon_1$&---& Target cell type 1 treatment efficacy\\
$k_1$& $\frac{\mbox{ml}}{\mbox{virions} \ \mbox{day}} $& Target cell type 1 infection rate\\
$\lambda_2$&$\frac{\mbox{cells}}{\mbox{ml} \ \mbox{day}}$& Target cell type 2 production rate\\
$d_2$&$\frac{1}{\mbox{day}}$& Target cell type 2 death rate\\
$f$&---& Treatment efficacy reduction in target cell type 2\\
$k_2$&$\frac{\mbox{ml}}{\mbox{virions} \ \mbox{day}}$& Target cell type 2 infection rate\\
$\delta$&$\frac{1}{\mbox{day}}$& Infected cell death rate\\
$m_1$&$\frac{\mbox{ml}}{\mbox{cells} \ \mbox{day}}$& Type 1 immune-induced clearance rate\\
$m_2$&$\frac{\mbox{ml}}{\mbox{cells} \ \mbox{day}}$& Type 2 immune-induced clearance rate\\
$\epsilon_2$&---& Target cell type 2 treatment efficacy\\
$N_T$&$\frac{\mbox{virions}}{\mbox{cell}}$& Virions produced per infected cell\\
$c$&$\frac{1}{\mbox{day}}$& Virus natural death rate\\
$\rho_1$&$\frac{\mbox{virions}}{\mbox{cell}}$& Average number of virions infecting a type 1 cell\\
$\rho_2$&$\frac{\mbox{virions}}{\mbox{cell}}$& Average number of virions infecting a type 2 cell\\
$\lambda_E$&$\frac{\mbox{cells}}{\mbox{ml} \ \mbox{day}}$& Immune effector production rate\\
$b_E$&$\frac{1}{\mbox{day}}$& Maximum birth rate for immune effectors\\
$K_b$&$\frac{\mbox{cells}}{\mbox{ml}}$& Saturation constant for immune effector birth\\
$d_E$&$\frac{1}{\mbox{day}}$& Maximum death rate for immune effectors\\
$K_d$&$\frac{\mbox{cells}}{\mbox{ml}}$& Saturation constant for immune effector death\\
$\delta_E$&$\frac{1}{\mbox{day}}$& Natural death rate for immune effectors\\		
		\end{tabular}
	\end{center}
	\label{tabpars}
\end{table}

\clearpage

As reported in \cite{adamsthesis,adams07}, data to be used with this model in inverse or parameter estimation problems typically consisted of monthly observations over a 3 year period (so approximately 36 longitudinal data points per patient) for the states $T_1+T_1^*$ and $V$. While this inverse problem is relatively ``small'' compared to many of those found in the literature, it still represents a nontrivial estimation challenge and is more than sufficient to illustrate the ideas and methodology we discuss in this presentation. Other difficult aspects (censored data requiring use of the Expectation Maximization algorithm as well as use of residual plots in attempts to validate the correctness of choice of corresponding statistical models introduced and discussed in the next section) of such inverse problems are discussed in the review chapter \cite{BDSS} and will not be pursued here.

%\clearpage
\section{Statistical Models for the Observation Process}
One has errors in any data collection process and the presence of this error is reflected in any parameter estimation results one might obtain. To understand and treat this, one usually specifies a {\em statistical model} for the observation process in addition to the {\em mathematical model} representing the dynamics. To illustrate ideas here we use ordinary least squares (OLS) consistent with an error model for absolute error in the observations. For a discussion of other frameworks (maximum likelihood in the case of known error distributions, generalized least squares appropriate for relative error models) see \cite{BDSS}.
Here the OLS estimation is based on the mathematical model for in-host HIV dynamics described above.  The observation process is formulated assuming there exists a vector $\theta_0\in\mathbb{R}^p$,
referred to as the {\em true parameter vector}, for which the model describes the log-scaled total number of
CD4$^{+}$ T-cells (uninfected and infected) exactly. It is also reasonably assumed that each of $n$ longitudinal observations $\{Y_i\}_{i=1}^{n}$
is affected by random deviations from the true underlying process. That is, if the mathematical model output is denoted by
\begin{equation}
	z(t_i;\theta_0)=\log_{10}\left(10^{x_1(t_i;\theta_0)}+10^{x_2(t_i;\theta_0)}\right),
\end{equation}	
then the statistical model for the scalar observation process is
\begin{equation}
	\begin{array}{lr}
	Y_i=z(t_i;\theta_0)+\mathcal{E}_i& \mbox{for } i=1,\dots,n.
	\end{array}
\end{equation}

The errors $\mathcal{E}_i$ are assumed to be random variables satisfying the following assumptions:
\begin{itemize}
	\item[(i)] the errors $\mathcal{E}_i$ have mean zero, $E[\mathcal{E}_i]=0$;
	\item[(ii)] the errors $\mathcal{E}_i$ have finite common variance, $\mbox{var}(\mathcal{E}_i)=\sigma_0^2<\infty$;
	\item[(iii)] the errors $\mathcal{E}_i$ are independent (i.e., $\mbox{cov}(\mathcal{E}_i,\mathcal{E}_j)=0$ whenever $i\neq j$) and identically distributed.
\end{itemize}
Assumptions (i)--(iii) imply that the mean of the observation is equal to the model output, $E[Y_i]=z(t_i;\theta_0)$,
and the variance in the observations is constant in time, $\mbox{var}(Y_i)=\sigma_0^2$.

The estimator $\theta_{OLS}=\theta_{OLS}^n$ minimizes
\begin{equation}\label{olscotfun}
	\sum_{i=1}^{n}[Y_i-z(t_i;\theta)]^2.
\end{equation}	
From  \cite{SeWi} we find that under a number of regularity and sampling conditions, as $n\rightarrow\infty$,
$\theta_{OLS}$ is approximately distributed according to a multivariate normal distribution, i.e.,
\begin{equation}
\theta_{OLS}^n\sim\mathcal{N}_p\left(\theta_0,\Sigma_0^n\right),
\end{equation}
where $\Sigma_0^n=\sigma^2_0[n\Omega_0]^{-1}
\in\mathbb{R}^{p\times p}$ and
\begin{equation}
    \Omega_0=\lim_{n\rightarrow\infty}\frac{1}{n} \chi^{n}(\theta_0)^{T}\chi^n(\theta_0).
\end{equation}
Asymptotic theory requires existence of this limit and non-singularity of $\Omega_0$.  The $p\times p$ matrix $\Sigma_0^n$
is the covariance matrix, and the $n\times p$ matrix $\chi^n(\theta_0)$ is known as the {\em sensitivity matrix} of the system, and is defined as
\begin{eqnarray} \label{defchimtrx}
    \chi_{ij}^n(\theta_0)=\left.\frac{\partial z(t_i;\theta)}{\partial \theta_j} \right|_{\theta=\theta_0} && 1\leq i\leq n, \ 1\leq j\leq p.
\end{eqnarray}
If $g\in\mathbb{R}^6$ denotes the right-side of Equations (\ref{x1eqn})--(\ref{x6eqn}), then
numerical values of $\chi^n(\theta)$ are readily calculated, for a particular $\theta$, by solving
 \begin{eqnarray}\label{seneqns}
 	\frac{dx}{dt}&=&g(t,x(t;\theta);\theta)\\
        \frac{d}{dt}\frac{\partial x}{\partial \theta}&=&\frac{\partial g}{\partial x}\frac{\partial x}{\partial \theta}+\frac{\partial g}{\partial \theta},
    \end{eqnarray}
from $t=t_0$ to $t=t_n$. One could alternatively solve for the sensitivity matrix using difference quotients (usually less accurately) or by using automatic differentiation software (for additional details on sensitivity matrix calculations see \cite{BDSS,BDE,ac08,ac09,eslami,finkAD}).

The estimate $\hat\theta_{OLS}=\hat\theta_{OLS}^n$ is a realization of the estimator $\theta_{OLS}$, and is calculated
using a realization $\{y_i\}_{i=1}^n$ of the observation process $\{Y_i\}_{i=1}^n$, while minimizing (\ref{olscotfun}) over $\theta$.
Moreover, the estimate $\hat\theta_{OLS}$ is used in the calculation
of the sampling distribution for the parameters.  The error variance $\sigma_0^2$ is approximated by
 \begin{equation}
    \hat\sigma^2_{OLS}=\frac{1}{n-p} \sum_{i=1}^{n}[y_i-z(t_i;\hat\theta_{OLS})]^2,
\end{equation}
while the covariance matrix $\Sigma_0^n$ is approximated by
\begin{equation}\label{covmtrx}
    \hat\Sigma_{OLS}^n= \hat\sigma^2_{OLS} \left[\chi(\hat\theta_{OLS}^n)^T\chi(\hat\theta_{OLS}^n)\right]^{-1}.
\end{equation}

As discussed in \cite{BDSS,DG,SeWi} an approximate for the sampling distribution of the estimator is given by
\begin{equation}
   \theta_{OLS}=\theta_{OLS}^n\sim \mathcal{N}_{p}(\theta_{0},\Sigma_{0}^n)\approx
    \mathcal{N}_{p}(\hat\theta_{OLS}^n,\hat\Sigma_{OLS}^n).
\end{equation}

Asymptotic standard errors can be used to quantify uncertainty in the estimation, and they are calculated by taking the square roots of the diagonal
elements of the covariance matrix $\hat\Sigma^n_{OLS}$, i.e.,
\begin{equation}\label{seeqn}
SE_k(\hat\theta_{OLS}^n)=\sqrt{(\hat\Sigma_{OLS}^n)_{kk}}, \quad
k=1,\dots,p.
\end{equation}

%\clearpage

\section{Subset Selection Algorithm}

The focus of our presentation here is how one chooses {\em a priori} (i.e., {\em before} any inverse problem calculations are carried out) which parameters and initial conditions can be readily estimated with a typical longitudinal data set. That is, from the parameters $\theta_2$ and initial conditions $\theta_1$, which components of $\theta=(\theta_1,\theta_2)$ yield a subset of readily identifiable parameters and initial conditions?
We illustrate an algorithm, developed recently in \cite{ac09}, to select parameter vectors that can be estimated from a given data set
using an ordinary least squares inverse problem formulation (similar ideas apply if one is using a relative error statistical model and generalized least squares formulations).  The algorithm searches all possible
parameter vectors and selects some of them based on two main criteria: (i) full rank of the sensitivity matrix, and (ii) uncertainty quantification
by means of asymptotic standard errors. Prior knowledge of a nominal set of values for all parameters along with the observation times for data (but not the values of the observations) will be required for our algorithm. Before describing the algorithm in detail and illustrating its use, we provide some motivation underlying the steps which involve the sensitivity matrix $\chi$ of \eqref{defchimtrx} and the Fisher Information Matrix $\mathcal{F}=\chi^T\chi$.

Ordinary least squares problems involve choosing $\Theta=\theta_{OLS}$ to minimize the difference between observations $Y$ and model output $z(\theta)$, i.e.,  minimize $|Y-z(\theta)|$ (here we use $|\cdot|$ for the Euclidean norm in $\mathbb{R}^n$). Replacing the the model with a first order linearization about $\theta_0$, we then wish to minimize
\[
|Y-z(\theta_0)-\nabla_{\theta}z(\theta_0)[\theta-\theta_0]|.
\]
If we use the statistical model $Y=z(\theta_0)+\mathcal{E}$ and let $\delta \theta =\theta-\theta_0$, we thus wish to minimize
\[
|\mathcal{E}-\chi(\theta_0)\delta\theta|,
\]
where $\chi=\nabla_{\theta}z$ is the $n\times p$ sensitivity matrix defined in \eqref{defchimtrx}. This is a standard optimization problem \cite[Section 6.11]{Lu} whose solution can be given using the pseudo inverse $\chi^{\dag}$ defined in terms of minimal norm solutions of the optimization problem and satisfying $\chi^{\dag}=(\chi^T\chi)^{\dag}\chi^T=\mathcal{F}^{\dag}\chi^T$. The solution is
\[
\delta \Theta=\chi^{\dag}\mathcal{E}
\]
or
\[
\Theta= \theta_0 + \chi^{\dag}\mathcal{E} = \theta_0 + \mathcal{F}^{\dag}\chi^T\mathcal{E}.
\]
If $\mathcal{F}$ is invertible, then the solution (to first order) of the OLS problem is
\begin{equation} \label{sol}
\Theta=\theta_0+ \mathcal{F}^{-1}\chi^T\mathcal{E}.
\end{equation}
From these calculations, we see that the rank of $\chi$ and the conditioning (or ill-conditioning) of $\mathcal{F}$ play a significant role in solving OLS inverse problems. Observe that the error (or noise) $\mathcal{E}$ in the data will in general be amplified as the ill-conditioning of $\mathcal{F}$ increases. We further note that the $n\times p$ sensitivity matrix $\chi$ is of full rank $p$ if and only if the $p\times p$ Fisher matrix $\mathcal{F}$ has rank $p$, or equivalently, is nonsingular. These underlying considerations have motivated a number of efforts (e.g., see \cite{BDE,BDEK,BEG}) on understanding the conditioning of the Fisher matrix as a function of the number $n$ and longitudinal locations  $\{t_i\}^n_{i=1}$ of data points as a key indicator for well-formulated inverse problems and as a tool in optimal design, especially with respect to computation of uncertainty (standard errors, confidence intervals) in parameter estimates.

Thus, we use an algorithm which first seeks sub-vectors of the parameter vector $\theta$ for which the corresponding sensitivity matrix has full rank and then use the normalized diagonals of the covariance matrix (the coefficients of variation) to rank the parameters among the resulting sub-vectors according to their potential for reliability in estimation.

In view of the comments above (which are very {\em local} in nature--both the sensitivity matrix and the Fisher Information Matrix are local quantities), one should be pessimistic about using these quantities to obtain any {\em nonlocal} selection methods or criteria for estimation. Indeed, for nonlinear complex systems, it is easy to argue that questions related to some type of global parameter identifiability are not fruitful questions to be pursuing.

As we have stated above, to apply the parameter subset selection algorithm we require prior knowledge of nominal variance and nominal
parameter values. These nominal values of $\sigma_0$ and $\theta_0$ are needed to calculate the sensitivity matrix, the Fisher matrix and the corresponding covariance matrix defined in \eqref{covmtrx}. For our illustration here, we use the variance and parameter estimates obtained in \cite{adamsthesis,adams07} for Patient \# 4 as nominal values. In problems for which no prior estimation has been carried out, one must use knowledge of the observation  process error and some knowledge of viable parameter values that might be reasonable with the model under investigation.

More precisely, here we assume the error variance is $\sigma_0^2= 1.100\times10^{-1}$, and assume the following nominal parameter values (for description and units see Table \ref{tabpars}):
\(
		x_1^0=\log_{10}(1.202\times 10^{3}),\ x_2^0=\log_{10}(6.165\times 10^{1}),\ x_3^0=\log_{10}(1.755\times 10^{1}),
		x_4^0=\log_{10}(6.096\times10^{-1}),\ x_5^0=\log_{10}(9.964\times 10^{5}),\ x_6^0=\log_{10}(1.883\times 10^{-1}),
		\lambda_1=4.633,\ d_1=4.533\times10^{-3},\ \epsilon_1= 6.017\times 10^{-1},
		k_1=1.976\times10^{-6},\ \lambda_2=1.001\times10^{-1},\ d_2=2.211\times10^{-2},
		f=5.3915\times10^{-1},\ k_2=5.529\times10^{-4},\ \delta=1.865\times10^{-1},
		m_1=2.439\times10^{-2},\ m_2=1.3099\times10^{-2},\ \epsilon_2=5.043\times10^{-1},
		N_T=1.904\times10^{1},\ c= 1.936\times10^{1},\ \rho_1=1.000,
		\rho_2=1.000,\ \lambda_E= 9.909\times10^{-3},\ b_E=9.785\times10^{-2},
		K_b=3.909\times10^{-1},\ d_E=1.021\times10^{-1},\ K_d= 8.379\times10^{-1}, \text{ and }
		\delta_E=7.030\times10^{-2}.
\)

\begin{figure}[h]
	\begin{center}
		\includegraphics[width=4.55in,height=2.5in]{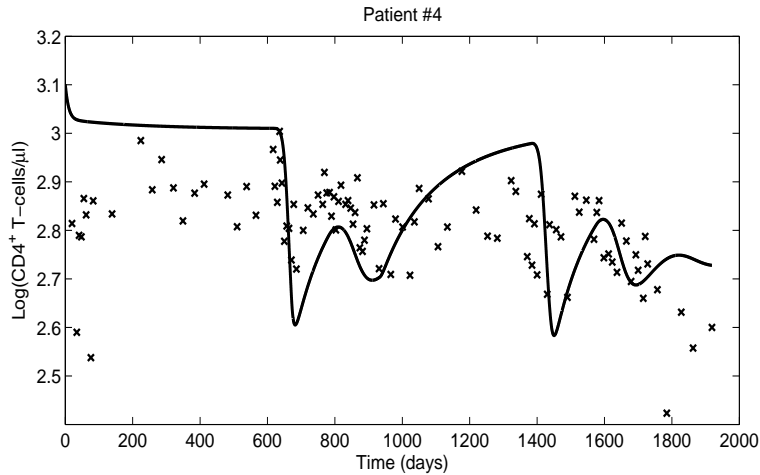}
	\end{center}
		\caption{Log-scaled data $\{y_i\}$ of Patient 4 CD4$^{+}$ T-cells (represented as `x'), and model output $z(t;\hat\theta_{OLS})$
 (represented by the solid curve) evaluated at
		parameter estimates obtained in \cite{adamsthesis,adams07}.}
	\label{pat4logdata}
\end{figure}

In Figure \ref{pat4logdata} we depict the log-scaled longitudinal observations (data) on the number of CD4$^{+}$ T-cells, $\{y_i\}$, and the model
output evaluated at the estimate (the nominal parameter values described above), $z(t_i;\hat\theta_{OLS})$, for Patient \#4 in \cite{adamsthesis,adams07}.

Given the vector
\[
	\theta=(\theta_1,\theta_2)\in\mathbb{R}^{28},
\]
for initial conditions plus system parameters,
we will consider sub-vectors, by partitioning into fixed and active (those to possibly be estimated) parameters.  It is assumed
the following entries are always fixed at known values provided in \cite{adamsthesis,adams07}:
$x_3^0$, $x_4^0$, $x_6^0$, $\rho_1$, and $\rho_2$.  In other words, we will calculate sub-vectors
from the $\mathbb{R}^{23}$ vector
\begin{equation}\label{q23}
	q=(x_1^0,x_2^0,x_5^0,\lambda_1,d_1,\epsilon_1,k_1,\lambda_2,d_2,f,k_2,\delta,m_1,m_2,
	\epsilon_2,N_T,c,\lambda_E,b_E,K_b,d_E,K_d,\delta_E ).
\end{equation}
For every fixed value of $p$, such that $p=2,3,\dots,22$, there are two partitions of interest: one with $p$ active parameters,
and the other one with $23-p$ fixed parameters.  For example, when $p=22$ one of twenty three possible partitions is the following:
fix $x_1^0$ and consider
\[
	(x_2^0,x_5^0,\lambda_1,d_1,\epsilon_1,k_1,\lambda_2,d_2,f,k_2,\delta,m_1,m_2,
	\epsilon_2,N_T,c,\lambda_E,b_E,K_b,d_E,K_d,\delta_E )^T\in\mathbb{R}^{22},
\]
as a vector with active parameters.  In the implementation of this subset selection algorithm, we carry out the calculation of all possible
vectors by using binary matrices with twenty eight columns,
such that every row has zeros for entries that are fixed, and ones for those that are active.  In the example above, the binary row is (recall that $x_3^0$, $x_4^0$, $x_6^0$, $\rho_1$, and $\rho_2$ are fixed throughout)
\[
	(0,1,0,0,1,0,1,1,1,1,1,1,1,1,1,1,1,1,1,1,0,0,1,1,1,1,1,1).
\]

For a fixed value of $p$ the set
\begin{equation}
	\mathcal{S}_p=\{\theta\in\mathbb{R}^p|\ \theta \mbox{ is a sub-vector of } q\in\mathbb{R}^{23} \mbox{defined in equation (\ref{q23})}\}
\end{equation}
collects all the possible active parameter vectors in $\mathbb{R}^p$.

%%%
We define the set
    \begin{equation}\label{viableq}
            \Theta_p=\{\theta|\ \theta\in \mathcal{S}_p \subset \mathbb{R}^{p},\ \mbox{rank}(\chi(\theta))=p\},
    \end{equation}
where $\chi(\theta)$ denotes the $n\times p$ sensitivity matrix.  By
construction, the elements of $\Theta_p$ are parameter vectors
that give sensitivity matrices with independent columns.

The next step in the selection procedure involves the
calculation of standard errors (uncertainty quantification) using
the asymptotic theory (see \eqref{seeqn}). For every
$\theta\in \Theta_p$, we define a vector of {\em coefficients of
variation} $\nu(\theta)\in \mathbb{R}^{p}$ such that for each
$i=1,\dots,p$,
        \[
            \nu_i(\theta)=\frac{\sqrt{(\Sigma(\theta))_{ii}}}{\theta_i},
        \]
        and
        \[
            \Sigma(\theta)=\sigma_0^2\left[\chi(\theta)^T\chi(\theta)\right]^{-1}\in\mathbb{R}^{p\times p}.
        \]

The components of the vector $\nu(\theta)$ are the
ratios of each standard error for a parameter to the corresponding
nominal parameter value. These ratios are dimensionless numbers
warrenting comparison even when parameters have considerably
different scales and units (e.g., $N_T$ is on the order of $10^1$, while $k_1$ is on the order of $10^{-6}$).  We then define
the {\em selection score} as
\[
        \alpha(\theta)=\left| \nu(\theta) \right|,
\]
where $|\cdot|$ is the norm in $\mathbb{R}^{p}$.
A selection score $\alpha(\theta)$ near zero indicates lower uncertainty
possibilities in the estimation, while large values of
$\alpha(\theta)$ suggest that one could expect to find substantial 
uncertainty in at least some of the components of the estimates in any parameter estimation attempt.

%%
%%%%%%

We summarize the steps of the algorithm as follows:
\begin{enumerate}
\item{\bf All possible active vectors.} For a fixed value of $p=2,\dots,22$,
fix $23-p$ parameters to nominal values, and then
calculate the set $\mathcal{S}_p$, which collects all the possible active parameter vectors in $\mathbb{R}^p$:
\[
	\mathcal{S}_p=\{\theta\in\mathbb{R}^p|\ \theta \mbox{ is a sub-vector of } q\in\mathbb{R}^{23} \mbox{defined in equation (\ref{q23})}\}.
\]

\item {\bf Full rank test}.
Calculate the set $\Theta_p$ as follows
    \[
    \Theta_p=\{\theta|\ \theta\in \mathcal{S}_p \subset \mathbb{R}^{p},\ \mbox{rank}(\chi(\theta))=p\}.
    \]
\item {\bf Standard error test.}  For every $\theta\in \Theta_p$
calculate a vector of coefficients of variation $\nu(\theta)\in
\mathbb{R}^{p}$ by
    \[
            \nu_i(\theta)=\frac{\sqrt{(\Sigma(\theta))_{ii}}}{\theta_i},
        \]
        for $i=1,\dots,p$, and
        \(
            \Sigma(\theta)=\sigma_0^2\left[\chi(\theta)^T\chi(\theta)\right]^{-1}\in\mathbb{R}^{p\times p}.
        \)
Calculate the selection score as
    \(
        \alpha(\theta)=\left|\nu(\theta) \right|.
    \)
    \end{enumerate}

%\clearpage

\section{Results and Discussion}

Results of the subset selection algorithm with the HIV model of Section \ref{modelsection} are given in Table \ref{tab5top}.  Parameter vectors, condition numbers (ratio of largest to smallest singular value \cite{golvan}),
and values of the selection score are displayed for $p=11$.  The third column of Table \ref{tab5top} displays selection score values from smallest (top) to largest (bottom).  For
the sake of clarity we only display five out of one million parameter vectors chosen by the selection algorithm.  The selection score values
range from $2.813\times10^{1}$ to $2.488\times10^{5}$ for the one million parameter vectors selected when $p=11$.

\begin{table}[h]
\caption{Parameter vectors obtained with subset selection algorithm for $p=11$.  For each
parameter vector $\theta\in\Theta_p$ the sensitivity matrix condition number $\kappa(\chi(\theta))$,
and the selection score $\alpha(\theta)$ are displayed.}
\begin{center}
\begin{tabular}{|c|c|c|} \hline\hline

Parameter vector, $\theta$ & Condition number, $\kappa(\chi(\theta))$ & Selection score, $\alpha(\theta)$ \\ \hline
$(x_1^0,x_5^0,\lambda_1,d_1,\epsilon_1,\lambda_2,d_2,k_2,\delta,\epsilon_2,N_T)$&3.083$\times 10^{5}$&2.881$\times 10^{1}$\\ \hline
$(x_1^0,x_5^0,\lambda_1,d_1,\epsilon_1,\lambda_2,d_2,k_2,\delta,\epsilon_2,c)$&3.083$\times 10^{5}$&2.884$\times 10^{1}$\\ \hline
$(x_1^0,x_5^0,\lambda_1,d_1,\epsilon_1,k_1,\lambda_2,d_2,k_2,\delta,\epsilon_2)$&2.084$\times 10^{8}$&2.897$\times 10^{1}$\\ \hline
$(x_2^0,x_5^0,\lambda_1,d_1,\epsilon_1,\lambda_2,d_2,k_2,\delta,\epsilon_2,N_T)$&2.986$\times 10^{5}$&2.905$\times 10^{1}$\\ \hline
$(x_2^0,x_5^0,\lambda_1,d_1,\epsilon_1,\lambda_2,d_2,k_2,\delta,\epsilon_2,c)$&2.986$\times 10^{5}$&2.907$\times 10^{1}$\\ \hline

\end{tabular}
\end{center}
\label{tab5top}
\end{table}

\begin{figure}
	\begin{center}
		\includegraphics[width=5in,height=3in]{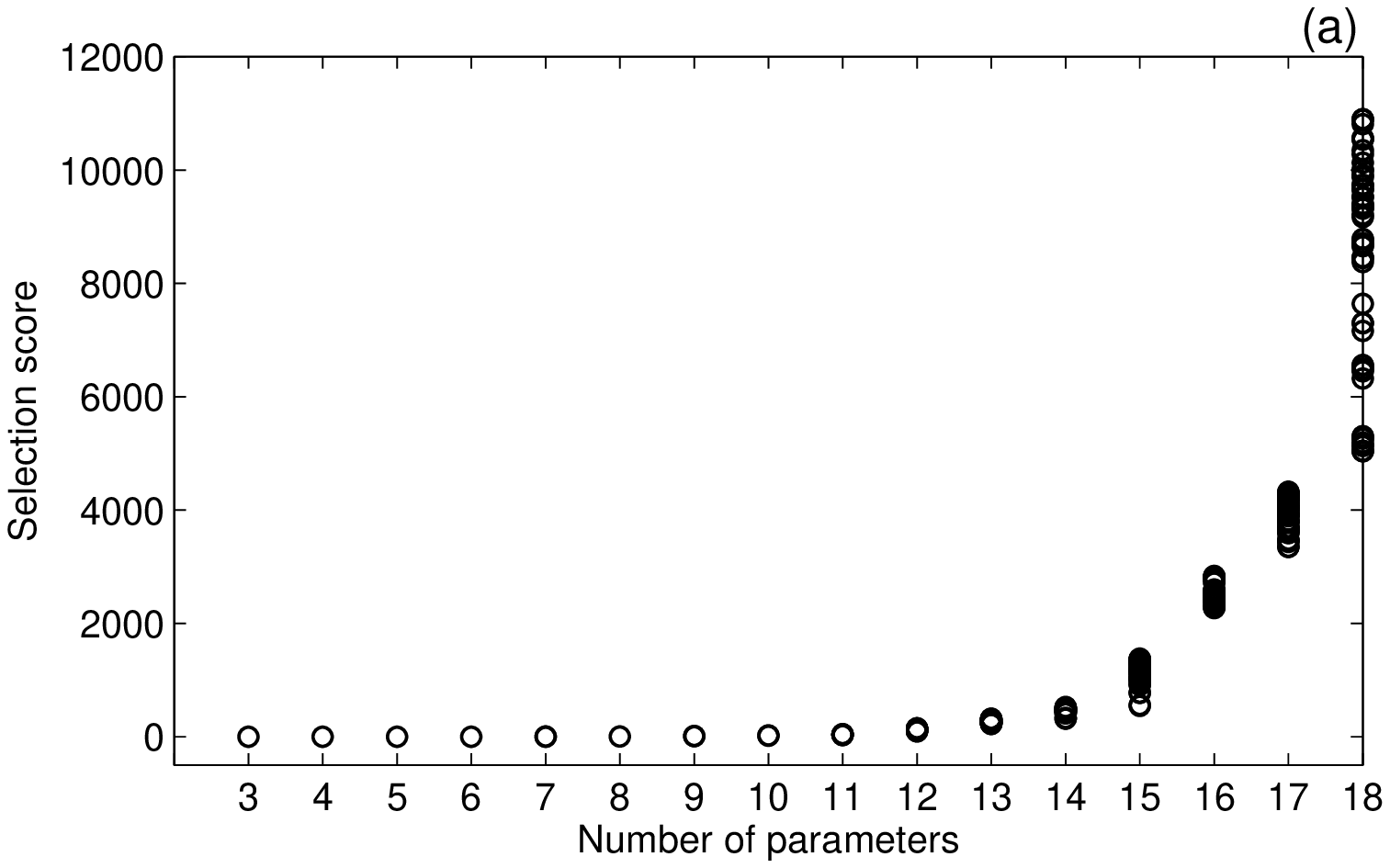}
		\includegraphics[width=5in,height=3in]{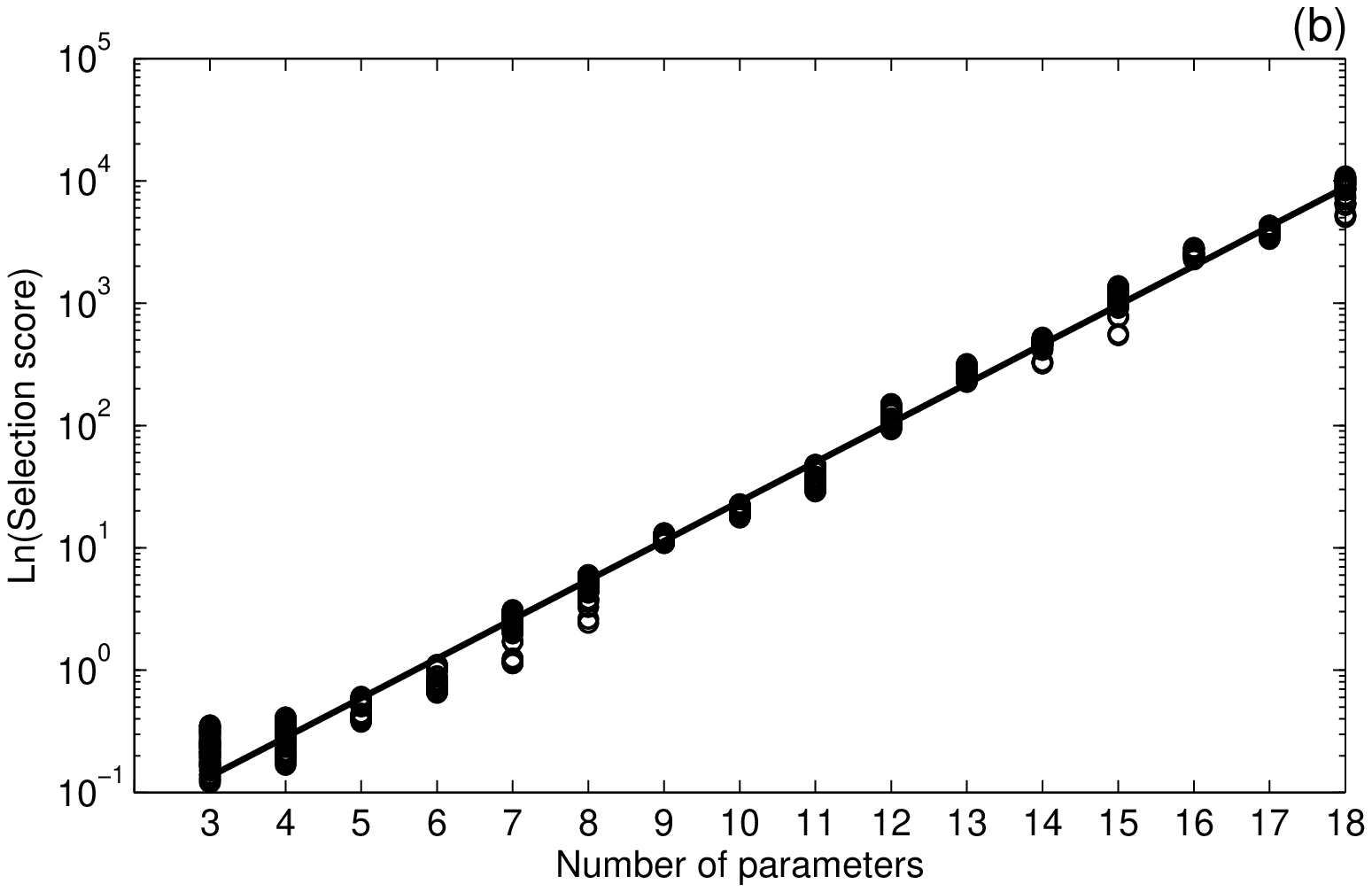}
	\end{center}
		\caption{(a) Selection score versus the number of parameters $p$. (b) Natural logarithm of selection score (circles) and regression line
		versus number of parameters $p$.  For each fixed value of $p$, the smallest 100 values of the selection score
		are displayed.}
	\label{selvsp}
\end{figure}

In \cite{adamsthesis,adams07}, the authors estimate the parameter vector
\[
	\theta=(x_1^0,x_2^0,x_5^0,\lambda_1,d_1,\epsilon_1,k_1,\epsilon_2,N_T,c,b_E) \in\mathbb{R}^{11}.
\]
The selection algorithm chooses most of these parameters.  For instance, the sub-vector $(x_5^0,\lambda_1,d_1,\epsilon_1,\epsilon_2)$
appears in every one of the top five parameter vectors displayed in Table \ref{tab5top}.  However, the sub-vector $(x_1^0,x_2^0,x_5^0)$ along with $b_E$
are never chosen among the top five parameter vectors. Even so, use of the subset selection algorithm discussed here (had it been available) might have proved valuable in the efforts reported in \cite{adamsthesis,adams07}.

In Figure \ref{selvsp}(a) we depict the selection score as a function of the number of parameters.  For each fixed value of $p$, one hundred values
are displayed, corresponding to the parameter vectors with the smallest one hundred selection score values.  Figure \ref{selvsp}(a) suggests that
parameter vectors with more than thirteen parameters ($13\leq p\leq 18$) might be expected to have large uncertainty when estimated
from observations, because the selection score ranges from $2.263\times 10^{2}$ to $1.090\times 10^{4}$.  Figure \ref{selvsp}(b) is a semilog
plot of Figure \ref{selvsp}(a), i.e., it displays the natural logarithm of the selection score as a function of the number of parameters.  Figure \ref{selvsp}(b)
also depicts the regression line, which fits the natural logarithm of the selection score.  From this linear regression we conclude the selection score $\alpha$
grows exponentially with the number of parameters to be estimated.  More precisely, for $3\leq p\leq18$, we find
\begin{equation}\label{alexp}
	\alpha\equiv\alpha(p)=Ce^{0.75p},
\end{equation}
where $C=8.52\times10^{-4}$.

\begin{figure}
	\begin{center}
		\includegraphics[width=5in,height=2.5in]{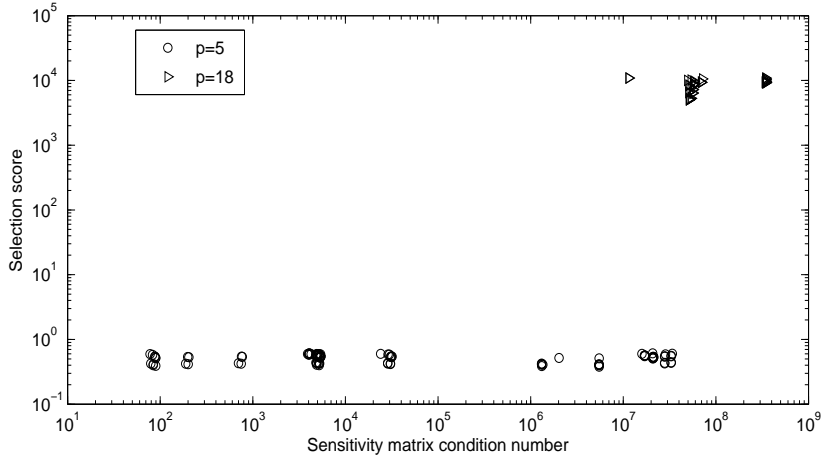}
	\end{center}
		\caption{Selection score $\alpha(\theta)$ versus condition number $\kappa(\chi(\theta))$, where $\theta\in\mathbb{R}^p$,
		for $p=5$ (circles) and $p=18$ (triangles).  Both axes are in logarithmic scale.  The smallest hundred values of the selection score
		are depicted for each value of $p$.}
	\label{selvscondp5p18}
\end{figure}

In Figure \ref{selvscondp5p18} we graph (in logarithmic scales) the smallest one hundred selection score values $\alpha(\theta)$ versus
the sensitivity matrix condition number $\kappa(\chi(\theta))$, with $\theta\in\mathbb{R}^p$, for $p=5$ (circles) and
$p=18$ (triangles).  The condition number $\kappa(\chi(\theta))$ is defined as the ratio of the largest to smallest singular value \cite{golvan} of the sensitivity matrix
$\chi(\theta)$.  It is clear from Figure \ref{selvscondp5p18} that the selection score drops dramatically from $p=18$ to $p=5$, which
is suggestive of a reduction in uncertainty quantification for these scenarios.  However, the conditioning of the sensitivity matrix does not exhibit
this decaying feature.  Some values of $\kappa(\chi(\theta))$ are within the same ball park, $10^{7}\leq\kappa(\chi(\theta))\leq10^{8}$
for $p=5$ and $p=18$, while other $\kappa(\chi(\theta))$ values for $p=5$ range considerably from  $7.768\times10^{1}$ to $5.486\times10^{6}$ .

%%>> min(condtoppars(find(condtoppars(:,1)<1e7),1))
%%   68.7139
%%>> max(condtoppars(find(condtoppars(:,1)<1e7),1))
%%   4.2920e+06

%\clearpage

In Table \ref{cvp5p18} we examine the effect that removing parameters from an estimation has in uncertainty quantification.  The coefficient of
variation (CV) is defined as the ratio of the standard error to the estimate for each parameter.  In Table \ref{cvp5p18} three cases are considered:
$p=18$, where $\theta=(x_1^0,x_2^0,x_5^0,\lambda_1,d_1,\epsilon_1,d_2,f,k_2,\delta,m_1,m_2,\epsilon_2,N_T,b_{E},K_{b},d_{E},K_{d})$;
$p=5$, where $\theta=(x_1^0,\lambda_1,\delta,\epsilon_2,N_{T})$; and $p=5$, where $\theta=(x_2^0,b_{E},K_{b},d_{E},K_{d})$.

There are consistent improvements in uncertainty quantification, 
with CV dropping as much as four orders of magnitude.  For instance, by comparing the second and third columns of
Table \ref{cvp5p18}, one sees the reduction of CV for $\lambda_1$, going from $8.430\times 10^{-1}$ to
$1.150\times 10^{-1}$, implies the standard error is 84\% of the estimate for $p=18$, while it reduces to 11\% of the estimate when $p=5$.    
For the parameter $N_{T}$, it is observed that the standard error reduces from being 40000\% to 10\% of the estimate.  A similar
remarkable improvement is also seen for $x_1^0$, with a standard error equal to 50000\% of the estimate for $p=18$, dropping to 
4\% of the estimate for $p=5$.  The improvement in uncertainty quantification is related to going from the 
upper right corner of Figure \ref{selvscondp5p18} into the lower left corner.  On one hand, the condition number and selection score for 
$\theta=(x_1^0,x_2^0,x_5^0,\lambda_1,d_1,\epsilon_1,d_2,f,k_2,\delta,m_1,m_2,\epsilon_2,N_T,b_{E},K_{b},d_{E},K_{d})$, 
are $7.518\times 10^{8}$ and $1.025\times 10^{5}$, respectively.  On the other hand, the condition number and selection score for
$\theta=(x_1^0,\lambda_1,\delta,\epsilon_2,N_{T})$ are $8.383\times 10^{1}$ and $3.990\times 10^{-1}$, respectively.

The fourth column of Table \ref{cvp5p18} is a reminder that reducing the number of parameters (e.g. from $p=18$ to $p=5$) is not enough
to guarantee reasonable improvements in uncertainty quantification.  Even though equation (\ref{alexp}) establishes an exponential relationship between
the norm of the vector of coefficients of variation and the number of parameters.  The best improvement in uncertainty quantification, while comparing the second
and fourth column of Table \ref{cvp5p18}, is observed for $x_2^0$, with a standard error equal to 2,000,000\% when $p=18$, which
drops to 200\% when $p=5$.  However, the latter is still an estimate with large uncertainty which must be avoided.

\begin{table}
\caption{Coefficient of variation (CV), defined as the ratio of standard error divided by estimate, for three parameter vectors.}
\begin{center}
\begin{tabular}{|c|c|c|c|}\hline
Parameter & CV ($p=18$) & CV ($p=5$) & CV($p=5$) \\ \hline
$x_1^0$&4.82$\times 10^{2}$ &4.10$\times 10^{-2}$ & ---\\ \hline
$x_2^0$&1.62$\times 10^{4}$  &---& 1.72$\times 10^{0}$\\ \hline
$x_5^0$&6.42$\times 10^{3}$   &---& ---\\ \hline
$\lambda_1$&8.43$\times 10^{-1}$   &1.15$\times 10^{-1}$ &--- \\ \hline
$d_1$&9.93$\times 10^{-1}$   &---& ---\\ \hline
$\epsilon_1$&1.24$\times 10^{2}$  &---& --- \\ \hline
$d_2$&3.79$\times 10^{1}$ &---&--- \\ \hline
$f$&4.94$\times 10^{2}$  &---& --- \\ \hline
$k_2$&4.70$\times 10^{1}$  &---&--- \\ \hline
$\delta$&3.98$\times 10^{2}$  &3.39$\times 10^{-1}$ & --- \\ \hline
$m_1$&2.24$\times 10^{4}$ &---&--- \\ \hline
$m_2$&3.82$\times 10^{4}$  &---&--- \\ \hline
$\epsilon_2$&2.06$\times 10^{2}$  &1.39$\times 10^{-1}$ & ---\\ \hline
$N_T$&4.04$\times 10^{2}$  &9.99$\times 10^{-2}$ & --- \\ \hline
$b_E$&6.10$\times 10^{4}$  &---&1.12$\times 10^{4}$ \\ \hline
$K_b$&2.51$\times 10^{4}$  &---&4.29$\times 10^{3}$ \\ \hline
$d_E$&5.79$\times 10^{4}$  &---&1.07$\times 10^{4}$ \\ \hline
$K_d$&2.30$\times 10^{4}$  &---&4.04$\times 10^{3}$ \\ \hline
\end{tabular}
\end{center}
\label{cvp5p18}
\end{table}

\clearpage
\section{Concluding Remarks}
As we have noted, inverse problems for complex system models containing a large number of parameters are difficult. There is great need for quantitative methods to assist in posing inverse problems that will be well formulated in the sense of the ability to provide parameter estimates with quantifiable small uncertainty estimates. We have introduced and illustrated use of such an algorithm that requires prior local information about ranges of admissible parameter values and initial values of interest along with information on the error in the observation process to be used with the inverse problem. These are needed in order to implement the sensitivity/Fisher matrix based algorithm.

Because sensitivity of a model with respect to a parameter is fundamentally related to the ability to estimate the parameter, and because sensitivity is a local concept, we observe that the pursuit of a global algorithm to use in formulating parameter estimation or inverse problems is most likely a quest that will go unfulfilled.

\section*{Acknowledgements} This research was
supported in part by Grant Number R01AI071915-07 from the National
Institute of Allergy and Infectious Diseases and in part by the
Air Force Office of Scientific Research under grant  number
FA9550-09-1-0226.  A. C.-A. carried portions of this work while visiting
the Statistical and Applied Mathematical Sciences Institute, which is funded
by the National Science Foundation under Grant DMS-0635449.  The content is solely the responsibility of the
authors and does not necessarily represent the official views of
the NIAID, the NIH, the AFOSR, or the NSF.

%This material was based upon work partially supported by the National Science Foundation under Grant DMS-0635449 
%to the Statistical and Applied Mathematical Sciences Institute. Any opinions, findings, and 
%conclusions or recommendations expressed in this material are those of the author(s) and do not necessarily 
%reflect the views of the National Science Foundation.

\end{document}